\documentclass[twocolumn,aps,superscriptaddress,showpacs,nofootinbib,floatfix]{revtex4-1}
\usepackage{graphicx} 
\usepackage{epsfig,bm}
\usepackage{graphicx,amssymb}
\usepackage{amsmath}
\usepackage{epstopdf}
\usepackage[eps]{pstricks}
\usepackage{slashed}
\def\Tr{{\rm{Tr}}}

\usepackage[normalem]{ulem}

\renewcommand{\sout}{\bgroup \color{red} \ULdepth=-.5ex \ULset}

\begin{document}

\title{Hadronic scattering effects on $\Lambda$ polarization in relativistic heavy ion collisions}

\author{Haesom Sung}
\email{ioussom@yonsei.ac.kr}
\affiliation{Department of Physics and Institute of Physics and Applied Physics, Yonsei University, Seoul 03722, Korea}
\affiliation{Cyclotron Institute, Texas $A\&M$ University, College Station, TX 77843, USA}

\author{Che Ming Ko}\email{ko@comp.tamu.edu}
\affiliation{Cyclotron Institute, Texas $A\&M$ University, College Station, TX 77843, USA}
\affiliation{Department of Physics and Astronomy, Texas $A\&M$ University, College Station, TX 77843, USA}

\author{Su Houng Lee}\email{suhoung@yonsei.ac.kr}
\affiliation{Department of Physics and Institute of Physics and Applied Physics, Yonsei University, Seoul 03722, Korea}

\begin{abstract}
The $\Lambda$ hyperon spin flip and non-flip cross sections are calculated in a simple hadronic model by including both the $s$-channel process involving the spin 3/2, positive parity $\Sigma^*(1358)$ resonance and the $t$-channel process via the exchange of a scalar $\sigma$ meson.  Because of its large mass, the $\Lambda$ spin flip to non-flip cross sections is negligibly small in the $t$-channel process compared to the constant value of 1/3.5 in the $s$-channel process. With the $s-$channel $\Lambda-\pi$ spin-dependent cross sections included in a schematic kinetic model, the effects of hadronic scatterings on the $\Lambda$ spin polarization in Au-Au collisions at $\sqrt{s_{NN}}=7.7$ GeV are studied. It is found that the $\Lambda$ spin polarization only decreases by 7-12\% during the hadronic stage of these collisions, which justifies the assumption in theoretical studies that compare the $\Lambda$ polarization calculated at the chemical freezeout to the measured one at the kinetic freezeout.  
\end{abstract}

\maketitle

\section{Introduction}

Lambda ($\Lambda$) hyperons produced in non-central Au-Au collisions at the Relativistic Heavy Ion Collider (RHIC) have been found by the STAR Collaboration~\cite{STAR:2017ckg} to be partially polarized in the direction perpendicular to the reaction plane, with a magnitude increasing with decreasing collision energy. The STAR Collaboration has also measured the dependence of $\Lambda$ local polarization along the beam direction on its azimuthal angle in the transverse plane of the collisions~\cite{Niida:2018hfw}.  Both observations can be described by the viscous hydrodynamic model~\cite{Fu:2021pok,Becattini:2021iol,Palermo:2024tza} that includes the contributions to the $\Lambda$ polarization from both the thermal vorticity field~\cite{Karpenko2017,Li:2017slc} and the thermal shear field~\cite{Liu:2021uhn,Becattini:2021suc,Yi:2021ryh} after assuming thermal equilibrium between the $\Lambda$ spin and the vortical fluid at the chemical freezeout of the collisions.   The experimental data can also be described by the non-equilibrium chiral kinetic approach~\cite{Sun:2016mvh,Sun:2018bjl} and a covariant angular-momentum-conserved chiral transport model~\cite{Liu:2019krs}, which automatically takes into account the effects of the vorticity and shear fields on quarks and antiquarks through their equations of motion and scatterings, by assuming that $\Lambda$ hyperons are formed from the coalescence of polarized $s$ quarks with polarized $u$ and $d$ quarks at hadronization~\cite{Sun:2017xhx}.  

Since both $\Lambda$ global and local spin polarizations decrease with temperature as the hadronic matter cools~\cite{Sun:2021nsg}, the fact that the experimental data is consistent with the $\Lambda$ spin polarization at the hadronization or chemical freezeout temperature indicates that it freezes out or decouples early from the hadronic matter. It was pointed out in Ref.~\cite{Ko:2023eyb} that this would be the case if the $\Lambda-\pi$ scattering is through the spin 3/2, parity positive $\Sigma^*(1358)$ resonance since the ratio of $\Lambda$ spin non-flip to flip probabilities in this scattering is 3.5 based on a simple consideration of $\Lambda$ spin and pion orbital angular momentum couplings in this scattering. A similar conclusion is reached in a recent study of the temperature dependence of the $\Lambda$ spin relaxation time in a hot pion gas based also on the $s$-channel $\Sigma^*(1358)$ resonance scattering~\cite{Hidaka:2023oze}. In the present study, the $\Lambda$ spin flip and non-flip cross sections by a pion are explicitly calculated not only for the $s$-channel process suggested in Ref.~\cite{Ko:2023eyb} but also for the $t$-channel process involving the exchange of the scalar $\sigma$ meson.  Including these cross sections in a schematic kinetic equation makes it possible to quantify the effect of hadronic scatterings on the $\Lambda$ spin polarization as the hadronic matter expands from the chemical to kinetic freezeout. Results from the present study shows that a spin polarized $\Lambda$ produced at the chemical freezeout of heavy ion collisions in Au-Au collisions at $\sqrt{s_{NN}}=7.7$ GeV only looses its spin polarization by about 7\% after it decouples at the kinetic freezeout.   This result thus justifies the assumption used in the literature of comparing the calculated $\Lambda$ spin polarization at chemical freezeout to the measured value in experiments. 

The paper is organized as follows.  In Sec. II, the $\Lambda-\pi$ spin-dependent scattering cross sections are calculated.  Their thermal averages are given in Sec. III, which is followed by the presentation of the schematic kinetic approach in Sec. IV.  The time evolution of the temperature of the hadronic matter and the pion fugacity in Au-Au collisions at $\sqrt{s_{NN}}=7.7$ GeV are given in Sec. V.  Results are then shown in Sec. VI and followed by  conclusions and discussions in Sec. VII.

\section{$\Lambda$ spin-dependent scattering cross sections with pion}

\begin{figure}[h] 
\centering
\includegraphics[scale=0.1]{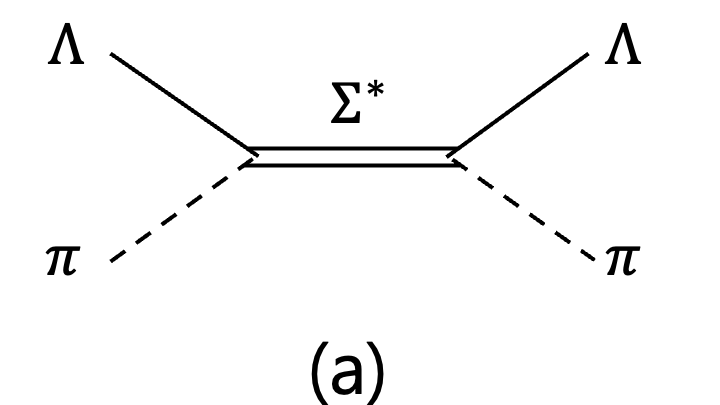}
\hspace{0.5cm}
\includegraphics[scale=0.1]{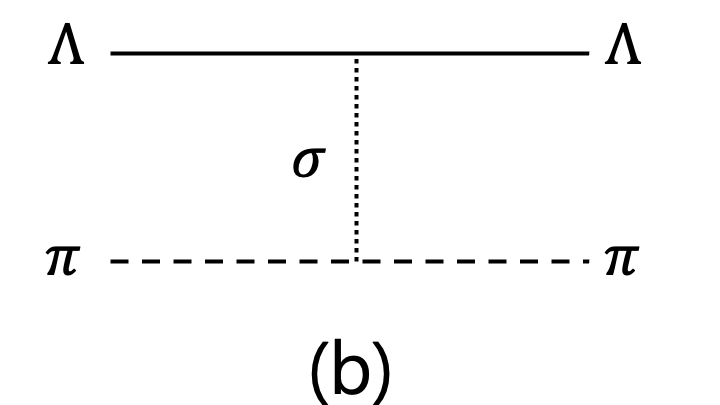}
\caption{Feynman diagrams for $\Lambda$ scattering by pion in (a) $s$-channel and (b) $t$-channel processes with solid and dashed lines indicating $\Lambda$ and $\pi$, respectively.  The double solid line in $s$-channel and the dotted line in $t$-channel denote $\Sigma^*(1358)$ resonance and $\sigma$ meson, respectively.} 
\label{Feynman}
\end{figure}

Since the hadronic matter produced in relativistic heavy ion collisions consists mostly of pions, only the spin-dependent scattering cross sections of $\Lambda$ with a pion are considered in the present study. The two Feynman diagrams for the $s$- and $t$-channel processes in this scattering are shown in Fig.~\ref{Feynman}.

\subsection{The $s$-channel process}

For the $s$-channel $\Lambda-\pi$ scattering through the $\Sigma^*(1358)$ resonance shown in  Fig.~\ref{Feynman}(a), the spin averaged cross section at center-of-mass energy $\sqrt{s}$ can be approximated by the Breit-Wigner formula, 
\begin{eqnarray}
    {\bar\sigma}(s)&=&\frac{2s_{\Sigma^*}+1}{(2s_{\Lambda}+1)(2s_{\pi}+1)}\frac{4\pi}{k^2}\frac{s\Gamma^2(s)}{(s-m_{\Sigma^*}^2)^2+s\Gamma^2(s)}\notag\\
    &=&\frac{8\pi}{k^2}\frac{s\Gamma^2(s)}{(s-m_{\Sigma^*}^2)^2+s\Gamma^2(s)},
    \label{eq:resonance}
\end{eqnarray}
where $s_\Lambda=1/2$, $s_\pi=0$, and $s_{\Sigma^*}=3/2$ are used for the spins of $\Lambda$, pion, and $\Sigma^*$, respectively. In Eq.(\ref{eq:resonance}), $m_{\Sigma^*}=1358$ MeV is the mass of $\Sigma^*$ resonance, $k$ is the magnitude of the momentum of $\Lambda$ and $\pi$ in their center-of-mass frame, i.e., 
\begin{equation}\label{mom}
k=\sqrt{\frac{[s-(m_\Lambda+m_\pi)^2][s-(m_\Lambda-m_\pi)^2]}{4s}},
\end{equation}
and $\Gamma(s)=\Gamma_{\Sigma^*}(k/k^*)^3$, with $\Gamma_{\Sigma^*}=36$ MeV being the width of $\Sigma^*$ at its peak mass~\cite{Olive:2016xmw} and $k^*$ given by Eq.(\ref{mom}) at $\sqrt{s}=m_{\Sigma^*}$. 

For a polarized $\Lambda$, its spin flip and non-flip scattering cross sections by a pion in the $s$-channel through the $\Sigma^*$ resonance are then proportional to the spin averaged scattering cross section in Eq.({\ref{eq:resonance}) with their relative magnitude determined by the Clebsch-Gordan coefficients $\langle j_\Lambda m_\Lambda l_\pi  m_\pi |J_{\Sigma^*}M_{\Sigma^*}\rangle$ for the coupling of the $\Lambda$ spin state $|j_\Lambda,m_\Lambda\rangle$, the pion orbital angular momentum state $|l_\pi,m_\pi\rangle$ to the $\Sigma^*$ spin state $|J_{\Sigma^*},M_{\Sigma^*}\rangle$.  For a $\Lambda$ having an initial spin pointing along the $z$-direction, the cross section for its spin to remain pointing in this direction after the scattering with a pion is 
\begin{eqnarray}
&&\sigma_{++}=\frac{A}{3}\sum_{m,M}\Big|\left\langle\frac{1}{2}\frac{1}{2}1m\Big|\frac{3}{2}M\right\rangle\Big|^4\bar\sigma\notag\\
&&\hspace{0.24in}=\frac{A}{3} \left[\Big|\left\langle \frac{1}{2}\frac{1}{2}11\Big|\frac{3}{2}\frac{3}{2}\right\rangle\Big|^4+\Big|\left\langle\frac{1}{2}\frac{1}{2}10\Big|\frac{3}{2}\frac{1}{2}\right\rangle\Big|^4\right.\notag\\
&&\hspace{0.36in}+\left.\Big|\left\langle\frac{1}{2}\frac{1}{2}1-1\Big|\frac{3}{2}-\frac{1}{2}\right\rangle\Big|^4\right]\bar\sigma\notag\\
&&\hspace{0.24in}=\frac{A}{3}\left[1+\left(\sqrt{\frac{2}{3}}\right)^4+\left(\sqrt{\frac{1}{3}}\right)^4\right]\bar\sigma=\frac{14A}{27}\bar\sigma,
\end{eqnarray}    
where the factor of 1/3 comes from averaging over the three components of the pion orbital angular momentum and $A$ is a normalization constant.  Similarly, the cross section for the $\Lambda$ to have its spin pointing in the negative $z$-direction after scattering with the pion is
\begin{eqnarray}
&&\sigma_{+-}=\frac{A}{3}\sum_{m,M}\Big|\left\langle\frac{1}{2}-\frac{1}{2}1m\Big|\frac{3}{2}M\right\rangle\left\langle\frac{3}{2}M\Big|\frac{1}{2}\frac{1}{2}1m\right\rangle\Big|^2\bar\sigma\notag\\
&&\hspace{0.24in}=\frac{A}{3}\left[\Big|\left\langle \frac{1}{2}-\frac{1}{2}11\Big|\frac{3}{2}\frac{1}{2}\right\rangle\left\langle\frac{3}{2}\frac{1}{2}\Big|\frac{1}{2}\frac{1}{2}10\right\rangle\Big|^2\right.\notag\\
&&\hspace{0.36in}+\left.\Big|\left\langle\frac{1}{2}-\frac{1}{2}10\Big|\frac{3}{2}-\frac{1}{2}\right\rangle\left\langle\frac{3}{2}-\frac{1}{2}\Big|\frac{1}{2}\frac{1}{2}1-1\right\rangle\Big|^2\right]\bar\sigma\notag\\
&&\hspace{0.24in}=\frac{A}{3}\left[\left(\sqrt{\frac{1}{3}}\sqrt{\frac{2}{3}}\right)^2+\left(\sqrt{\frac{2}{3}}\sqrt{\frac{1}{3}}\right)^2\right]\bar\sigma=\frac{4A}{27}\bar\sigma.
\end{eqnarray}    
The ratio of the spin non-flip to spin flip cross sections of a polarized $\Lambda$ is thus $R=\sigma_{++}/\sigma_{+-}= 3.5$ as given in Ref.~\cite{Ko:2023eyb}. Requiring $\sigma_{++}+\sigma_{+-}=\frac{2A}{3}\bar\sigma=\bar\sigma$ gives $A=3/2$, which leade to $\sigma_{++}=\frac{7}{9}\bar\sigma$ and $\sigma_{+-}=\frac{2}{9}\bar\sigma$. Shown in Fig.~\ref{cross} by the solid line is the $s$-channel spin-averaged $\Lambda-\pi$ resonance scattering cross section as a function of  $\sqrt{s}-\sqrt{s_0}$ with $\sqrt{s_0}=m_{\Lambda}+m_\pi$, which is seen to peak at $\sqrt{s}=m_{\Sigma^*}$ as expected. 

\begin{figure}[h]
\centering
\includegraphics[width=0.9\linewidth]{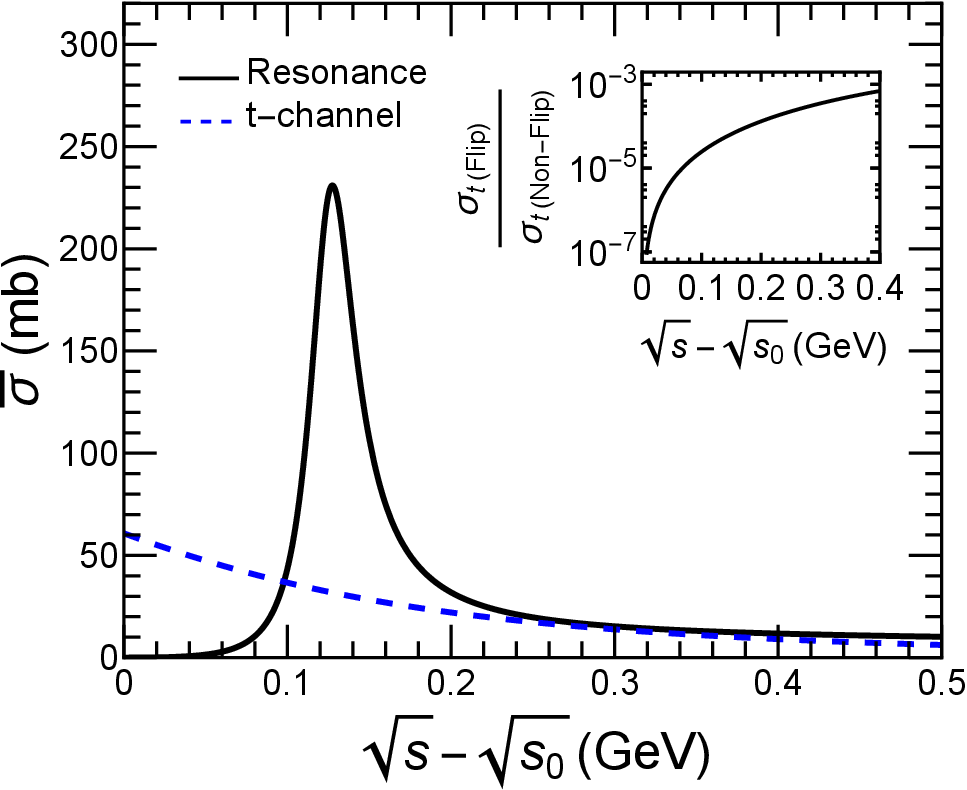}
\caption{Spin averaged $\Lambda +\pi\rightarrow \Lambda +\pi$ cross section for the $s$-channel (solid line) and the $t$-channel (dashed line) process as functions of $\sqrt{s}-\sqrt{s_0}$. Also shown in the inset is the ratio of spin flip to non-flip cross sections for the $t$-channel process.}
\label{cross}
\end{figure}

\subsection{The $t$-channel process}

The $t$-channel contribution to $\Lambda-\pi$ scattering shown by the Feynman diagram in Fig.~\ref{Feynman}(b) can be evaluated by using the following interaction Lagrangian densities,
\begin{eqnarray}
&&{\cal L}_{\Lambda\Lambda\sigma}=g_{\Lambda\Lambda\sigma}\bar\Lambda\Lambda\sigma,\notag\\
&&{\cal L}_{\sigma\pi\pi}=g_{\sigma\pi\pi}\sigma\pi\pi. 
\end{eqnarray}   
For the coupling constant $g_{\Lambda\Lambda\sigma}$, it has the value $g_{\Lambda\Lambda\sigma}=\frac{2}{3}g_{NN\sigma}=7.07$ if one uses the $g_{NN\sigma}^2/4\pi=8.94$ in the relativistic one-boson exchange nucleon-nucleon potential Ref.~\cite{Machleidt:1989tm} and the light quark counting rule for the relation between the couplings of light baryons and strange baryons to light mesons.  The $g_{\pi\pi\sigma}$ can be determined from the $\sigma$ meson decay width according to 
\begin{equation}
    \Gamma_{\sigma}= \frac{g^2_{\sigma\pi\pi}}{8\pi m^2_\sigma}\frac{\sqrt{m^2_\sigma-4m_{\pi}^2}}{2}.
\end{equation}
Using $m_\sigma\approx 550$ MeV~\cite{Machleidt:1989tm} and $\Gamma_{\sigma}(s=m_\sigma^2)\approx 240$ MeV~\cite{Olive:2016xmw}, the coupling constant $g_{\sigma\pi\pi}$ is found to have the value $g_{\sigma\pi\pi}= 2.77$ GeV.

For the $t$-channel process in $\Lambda(p_1)+\pi(p_2)\rightarrow \Lambda(p_3)+\pi(p_4)$ scattering, its invariant amplitude is 
\begin{equation}
    i\mathcal{M}_{t} = g_{\Lambda\Lambda\sigma}g_{\sigma\pi\pi}\bar{u}(p_3)\frac{i}{t-m_{\sigma}^2+im_{\sigma}\Gamma_{\sigma}}u(p_1), 
\end{equation}
where $t=(p_1-p_3)^2=-2p^2(1-\cos\theta)$ with $p$ and $\theta$ being the magnitude of and angle between $\Lambda$ initial and final momenta in their center-of-mass frame, respectively.  The spin-averaged squared invariant amplitude for the $t$-channel process is then  
\begin{eqnarray}\label{eq:Mt}
   |\overline{\mathcal{M}_t}|^2 &=& \frac{g_{\Lambda\Lambda\sigma}^2g_{\sigma\pi\pi}^2}{2}\frac{\Tr\left[(\slashed{p}_3 +m_{\Lambda})(\slashed{p}_{1}+m_{\Lambda})\right]}{(t-m_{\sigma}^2)^2+m_{\sigma}^2\Gamma^2_{\sigma}}
    \notag \\
    &=& 2 g_{\Lambda\Lambda\sigma}^2g_{\sigma\pi\pi}^2 \frac{p_1\cdot p_3 +m_{\Lambda}^2}{(t-m_{\sigma}^2)^2+m_{\sigma}^2\Gamma^2_{\sigma}}
    \notag \\
    &=& g_{\Lambda\Lambda\sigma}^2g_{\sigma\pi\pi}^2 \frac{4m_{\Lambda}^2-t}{(t-m_{\sigma}^2)^2+m_{\sigma}^2\Gamma^2_{\sigma}}. 
\end{eqnarray}

To take into account the finite size of hadrons, form factors are introduced at the $\Lambda\Lambda\sigma$ and $\sigma\pi\pi$ vertices. Using the monopole form factor of the form 
\begin{equation}
F(t)=\frac{\Lambda^2-m_\sigma^2}{\Lambda^2-t},
\end{equation}
with the value $\Lambda=1.9$ GeV as in Ref.~\cite{Machleidt:1989tm}, the spin-averaged differential $t$-channel cross section can then be calculated from 
\begin{equation}\label{eq:cross}
    \frac{d\sigma}{d\Omega}=\frac{1}{64\pi^2s}|\overline{\mathcal{M}_t}|^2F^4(t).
\end{equation}
The total spin-averaged $t$-channel cross section after integrating the scattering angles is shown by the dashed line in Fig.~\ref{cross} as a function of $\sqrt{s}-\sqrt{s_0}$. Its value decreases with increasing $\sqrt{s}$ and is about a factor of 8 smaller than the peak $s$-channel cross section at energy near the $\Sigma^*$ resonance.  

Although it is possible to calculate the $\Lambda$ spin flip and non-flip cross sections in the $t$-channel process by using the $\Lambda$ spin projection operator, the explicit spin-dependent spinors of the $\Lambda$ are used in the present study.  In this approach, the spin-dependent amplitude is proportional to the scalar product of $\Lambda$ spinors with initial spin direction $s_i$ and final spin direction $s_f$, i.e., 
\begin{eqnarray}\label{amplitude}
&&    \bar{u}_{s_f}(p_3)u_{s_i}(p_1) = (E_{\Lambda}+m_{\Lambda})
    \begin{pmatrix}\label{amplidute}
        \chi^{\dagger}_{s_f}, \chi^{\dagger}_{s_f}\frac{{\bf p}_3\cdot{\boldsymbol\sigma}}{E_{\Lambda}+m}
    \end{pmatrix}
    \notag \\
&&\hspace{1in}\times
    \begin{pmatrix}
        \chi_{s_i}
        \\
        -\frac{{\bf p}_1\cdot{\boldsymbol\sigma}}{E_{\Lambda}+m_{\Lambda}}\chi_{s_i}
    \end{pmatrix}
    \notag \\
&&\hspace{0.5in}=(E_{\Lambda}+m_{\Lambda})\left\{ \left[1-\frac{{\bf p}_3\cdot{\bf p}_1}{(E_{\Lambda}+m_{\Lambda})^2}\right]\chi^{\dagger}_{s_{f}}\chi_{s_{i}}\right.
    \notag\\
&&\hspace{0.6in}\left.-i\frac{{\bf p}_{3}\times{\bf p}_{1}}{(E_{\Lambda}+m_{\Lambda})^2}\cdot\chi^{\dagger}_{s_{f}}{\boldsymbol\sigma}\chi_{s_{i}}\right\}.
\end{eqnarray}

The $\Lambda$ spin flip and non-flip amplitudes can be explicitly evaluated from Eq.(\ref{amplitude}) by taking the $x-z$ plane to be the scattering plane with the $\Lambda$ initial momentum ${\bf p}_1=p\hat z$ and final momentum ${\bf p}_3=p(\sin\theta\hat x+\cos\theta\hat z)$, which leads to ${\bf p}_3\times{\bf p}_1=-p^2\sin\theta\hat y$ and thus ${\bf p}_3\times{\bf p}_1\cdot\chi^\dagger_{s_f}{\boldsymbol\sigma}\chi_{s_i}=-p^2\sin\theta\chi^\dagger_{s_f}\sigma_y\chi_{s_i}$.   Along a direction characterized by the polar angles $(\theta_s,\phi_s)$, the $\Lambda$ spin up and down states are 
\begin{eqnarray}
\chi_+=\left(\begin{matrix}\cos\frac{\theta_s}{2}\cr e^{i\phi_s}\sin\frac{\theta_s}{2}\end{matrix}\right),~
\chi_-=\left(\begin{matrix}\sin\frac{\theta_s}{2}\cr -e^{i\phi_s}\cos\frac{\theta_s}{2}\end{matrix}\right).
\end{eqnarray}
Using these expressions and $\sigma_y=\left(\begin{matrix}0&-i\cr i&0\end{matrix}\right)$, one has 
\begin{eqnarray}
\chi^\dagger_+\sigma_y\chi_+ &=&\sin \phi_s \sin\theta_s,\notag\\
\chi^\dagger_-\sigma_y\chi_+ &=&-\cos\theta_s\sin\phi_s-i\cos\phi_s,
\end{eqnarray}
which lead to 
\begin{eqnarray}
&&\bar{u}_{+}(p_3)u_{+}(p_1) = (E_{\Lambda}+m_{\Lambda})\Bigg[1-\frac{p^2\cos\theta}{(E_\Lambda+m_\Lambda)^2}\notag \\
&&\hspace{1 in} +i\frac{p^2\sin\theta}{\left(E+m\right)^2}\sin \phi_s \sin\theta_s\Bigg],\\
&&\bar{u}_{-}(p_3)u_{+}(p_1) =\frac{p^2\sin\theta}{E_\Lambda+m_\Lambda}(\cos\phi_s-i\cos\theta_s\sin\phi_s).\notag
\end{eqnarray}
With above results, one obtains the following expression for the squared $\Lambda$ spin non-flip and flip amplitudes after including the coupling constants from the vertices and the propagator of exchanged $\sigma$ meson,
\begin{eqnarray}
&&|\mathcal{M}_{t++}|^2=\frac{g_{\Lambda\Lambda\sigma}^2g_{\sigma\pi\pi}^2}
{(t-m_{\sigma}^2)^2+m_{\sigma}^2\Gamma_{\sigma}^2}\notag\\ \label{eq:Mtuu}
&&\hspace{0.7in}\times(E_\Lambda+m_\Lambda)^2\Bigg[\left(1-\frac{p^2\cos\theta}{(E_\Lambda+m_\Lambda)^2}\right)^2
\notag \\
&&\hspace{0.7in} +\frac{p^4\sin^2\theta}{\left(E+m\right)^4}\sin^2{\theta_s} \sin^2{\phi_s}\Bigg], \\
&&|\mathcal{M}_{t+-}|^2=\frac{g_{\Lambda\Lambda\sigma}^2g_{\sigma\pi\pi}^2}
{(t-m_{\sigma}^2)^2+m_{\sigma}^2\Gamma_{\sigma}^2}\frac{p^4\sin^2\theta}{(E_\Lambda+m_\Lambda)^2}
\notag \\
&&\hspace{0.7in}\times(\cos^2\theta_s\sin\phi^2_s+\cos^2\phi_s). \label{eq:Mtud}
\end{eqnarray}
It can be easily shown from Eqs.(\ref{eq:Mt}), (\ref{eq:Mtuu}), and (\ref{eq:Mtud}) that $|\mathcal{M}_{t++}|^2+|\mathcal{M}_{t+-}|^2=|\overline{\mathcal{M}_t}|^2$ as expected.  

Eq.(\ref{eq:Mtud}) shows that, with the $\Lambda$ having initial momentum in the $z$-direction and final momentum in the $x-z$ plane, it has the largest spin-flip cross section if it is initially longitudinally polarized along the $z$-direction ($\theta_s=0$) or transversely polarized along the $x$-direction ($\theta_s=\pi/2$ with $\phi_s=0$). In these cases, Eqs.(\ref{eq:Mtuu}) and (\ref{eq:Mtud}) become
 
\begin{eqnarray}
&&|\mathcal{M}_{t++}|^2=\frac{g_{\Lambda\Lambda\sigma}^2g_{\sigma\pi\pi}^2}
{(t-m_{\sigma}^2)^2+m_{\sigma}^2\Gamma_{\sigma}^2}\notag\\ \label{eq:Mtuu1}
&&\hspace{0.7in}\times(E_\Lambda+m_\Lambda)^2\left[1-\frac{p^2\cos\theta}{(E_\Lambda+m_\Lambda)^2}\right]^2,\\
&&|\mathcal{M}_{t+-}|^2=\frac{g_{\Lambda\Lambda\sigma}^2g_{\sigma\pi\pi}^2}
{(t-m_{\sigma}^2)^2+m_{\sigma}^2\Gamma_{\sigma}^2}\frac{p^4\sin^2\theta}{(E_\Lambda+m_\Lambda)^2}. \label{eq:Mtud1}
\end{eqnarray}
The ratio of spin flip to non-flip cross sections calculated from Eq.(\ref{eq:cross}) using Eqs.(\ref{eq:Mtuu1}) and (\ref{eq:Mtud1}) and after integrating the scattering angle is shown in the inset of Fig.~\ref{cross}, which is seen to be negligibly small even for the case that the polarized $\Lambda$ has the largest spin-flip cross section.  This result is not surprising because the flip of $\Lambda$ spin due to the exchange of a scalar meson is a relativistic effect, which is small because of the small $\Lambda$ velocity in the hadronic matter produced in relativistic heavy ion collisions as a result of its large mass.  

\section{Thermally averaged cross sections}

\begin{figure}[h]
\centering
\includegraphics[width=0.9\linewidth]{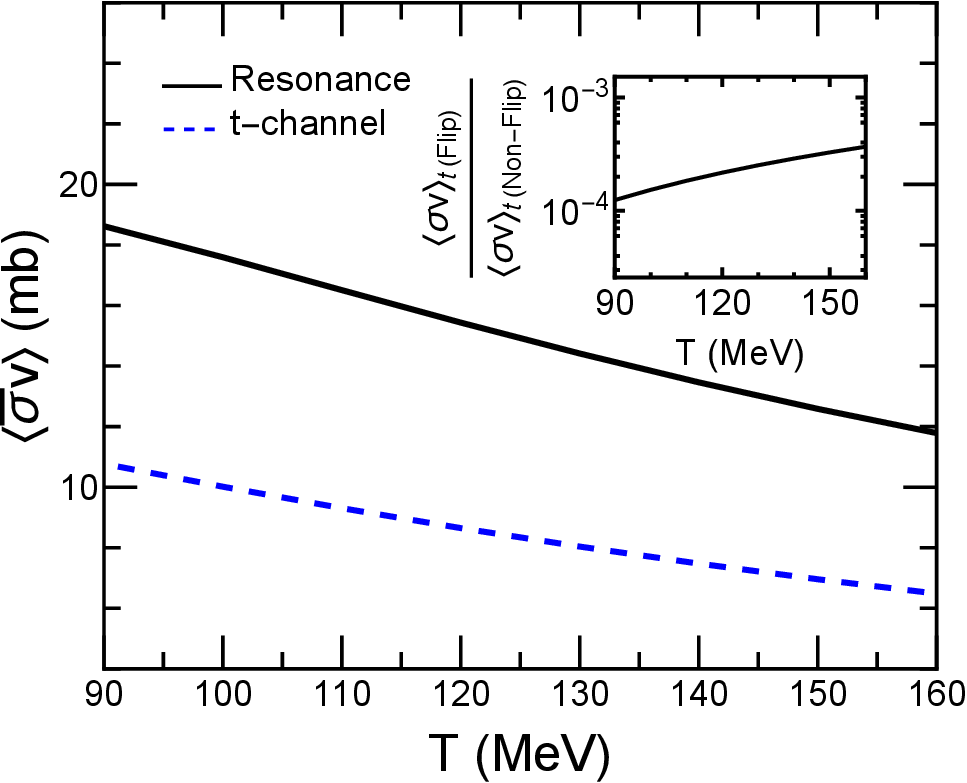}
\caption{Thermally averaged $\Lambda-\pi$ scatting cross section in $s$-channel (solid line) and $t$-channel (dashed line)
as functions of temperature.  Also shown in the inset is the ratio of thermally averaged $t-$channel $\Lambda$ spin flip to non-flip cross sections as a function of temperature.}
\label{tcross}
\end{figure}

For $\Lambda$ scattering in a thermalized hadronic matter, the time evolution of its spin polarization in the kinetic approach depends on the thermally averaged $\Lambda-\pi$ scattering cross sections, 
\begin{eqnarray}\label{eq:AvgCrs}
&&\langle\sigma_{ab\rightarrow cd}v_{ab}\rangle =\frac{1}{1+\delta_{ab}} \frac{\int d^{3}{\bf p}_{a}d^{3}{\bf p}_{b}f_{a}({\bf p}_{a})f_{b}({\bf p}_{b})\sigma_{ab\rightarrow cd}v_{ab}}{\int d^{3}{\bf p}_{a}d^{3}{\bf p}_{b}f_{a}({\bf p}_{a})f_{b}({\bf p}_{b})}\notag\\
&&\hspace{0.2in}=\frac{1}{1+\delta_{ab}}\frac{T^4}{4m_a^2K_2(m_a/t)m_b^2K_2(m_b/T)}\notag\\
&&\hspace{0.3in}\times\int_{z_0}^\infty da K_1(z)\sigma_{ab\to cd}(z^2T^2)[z^2-(m_a+m_b)^2/T^2]\notag\\
&&\hspace{0.3in}\times[z^2-(m_a-m_b)^2/T^2],
\end{eqnarray}
In the above, $f_{i}({\bf p}_{i})$ is the Boltzmann momentum distribution of particle species $i=a, b$, i.e., $f_i({\bf p}_{i})=e^{-\sqrt{{\bf p}_i^{2}+m_i^{2}}/T}$ with $m_i$ being the particle mass, and $v_{ab}$ is the relative velocity between particles $a$ and $b$. The functions $K_1(z)$ and $K_2(m_{a,b}/T)$ are the modified Bessel functions of first and second kinds, respectively. Figure~\ref{tcross} shows the thermally averaged $s$- and $t$-channel spin-averaged scattering cross sections of $\Lambda$ by a pion. It is seen that the one from the $t$-channel process is only about a factor of 1.7 smaller than that from the $s$-channel process at all temperature of the hadronic matter, indicating that the average invariant mass of a $\Lambda$ and $\pi$ pair in the hadronic matter is below the $\Sigma^*$ mass. Also shown in the inset is the ratio of thermally averaged $t-$ channel $\Lambda$ spin flip to non-flip cross sections, which is again negligibly small as in the case for their spin-averaged cross sections. The effect of the $t-$channel $\Lambda-\pi$ scattering process on the $\Lambda$ spin polarization will thus be neglected in the present study. 

\section{Kinetic equations for $\Lambda$ spin evolution} 
 
Using the thermally averaged $s-$channel $\Lambda-\pi$ scattering cross sections in the previous section, the numbers of spin-up and spin-down $\Lambda$ hyperons, denoted by  $N_{\Lambda_+}$ and $N_{\Lambda_-}$, respectively, in the kinetic approach change with time according to the following equations, 
\begin{eqnarray}
&&\frac{dN_{\Lambda_+}}{d\tau}=-\langle\sigma_{\Lambda_{+}\pi\rightarrow\Lambda_{-}\pi}v\rangle z_\pi n^{(0)}_{\pi}N_{\Lambda_{+}}\notag\\
&&\hspace{0.55in}+\langle\sigma_{\Lambda_{-}\pi\rightarrow\Lambda_{+}\pi}v\rangle z_\pi n^{(0)}_{\pi}N_{\Lambda_{-}},
\label{Eq:kinetic1}\\
&&\frac{dN_{\Lambda_{-}}}{d\tau} = \langle\sigma_{\Lambda_{+}\pi\rightarrow\Lambda_{-}\pi}v\rangle z_\pi n^{(0)}_{\pi}N_{\Lambda_{+}}\notag\\
&&\hspace{0.55in}-\langle\sigma_{\Lambda_{-}\pi\rightarrow\Lambda_{+}\pi}v\rangle z_\pi n^{(0)}_{\pi}N_{\Lambda_{-}}\label{Eq:kinetic2}.
\end{eqnarray}
In the above, $n^{(0)}_{\pi}$ is the density of thermally equilibrated pions, given by $n_\pi^{(0)} =g_\pi\int\frac{d^3{\bf p}}{(2\pi)^3}f_\pi(p)$ with $g_\pi=3$ due to its isospin degeneracy, and $z_\pi$ is its fugacity to take into account the fact that the effective pion number, which includes both free pions and those from resonance decays, remains unchanged during the expansion and cooling of the hadronic matter according to the statistical hadronization model~\cite{Andronic:2005yp}. 

Using the relation $\langle\sigma_{\Lambda_{+}\pi\rightarrow\Lambda_{-}\pi}v\rangle= \langle\sigma_{\Lambda_{-}\pi\rightarrow\Lambda_{+}\pi}v\rangle\equiv\langle\sigma v\rangle$, the number difference between spin-up and spin-down $\Lambda$ hyperons, $\Delta N(\tau) = N_{\Lambda_{+}}(\tau)-N_{\Lambda_{-}}(\tau)$, then satisfies the equation,}
\begin{equation}
\frac{d\Delta N}{d\tau}=-2\langle\sigma v\rangle z_\pi n^{(0)}_\pi\Delta N,
\end{equation}
which can be solved to give
\begin{equation}\label{eq:number}
\Delta N(\tau) = \Delta N_0 e^{-2\int \left<\sigma v\right>z_\pi n^{(0)}_{\pi} d\tau},
\end{equation}
where $\Delta N_0$ denotes the initial number difference between spin up and spin down $\Lambda$ hyperons.

\section{Time evolution of temperature and pion fugacity in hadronic matter}

Since the largest $\Lambda$ polarization reported by the STAR Collaboration at beam energy scan experiments at RHIC is from Au-Au central 20-50\% collisions at $\sqrt{s_{NN}}=7.7$ GeV~\cite{STAR:2017ckg}, the hadronic scattering effects on $\Lambda$ polarization in these collisions are considered in the present study.  According to the statistical hadronization model based on the grand canonical ensemble fit to the experimentally measured yields of produced particles in 30-40\% collision centrality of these collisions Ref.~\cite{STAR:2017sal}, the extracted chemical freezeout temperature, baryon chemical potential, and strangeness chemical potential are about $T_{\rm ch}=146$ MeV, $\mu_B=376$ MeV, and $\mu_S=88$ MeV, respectively. Also, using the blast-wave model to fit the transverse momentum spectra of measured particles, a kinetic freezeout temperature of $T_{\rm k}=129$ MeV is found in Ref.~\cite{STAR:2017sal}.   The small charged chemical potential given in Ref.~\cite{STAR:2017sal} is, however, neglected in the present study.

\begin{table}[h]
    \centering
    \begin{tabular}{ c c c c c  }
    \hline
         $\sqrt{s_{NN}}$ &$T_{\rm ch}$ & $T_{\rm k}$ & $\tau_{\rm ch}$ & $\tau_{\rm k}$  \\
         (GeV)& (MeV) & (MeV) & (fm/c) & (fm/c) \\
    [0.5ex]
    \hline
         7.7 & 146 & 129 & 5.13 & 6.46 \\
    \hline
    \end{tabular}
    \caption{Parameter values in Eq.(\ref{Temp_eq}) for the time evolution of the temperature of the hadronic matter produced in Au-Au collisions at $\sqrt{s_{NN}}=7.7$ GeV.}
    \label{table:temperature}
\end{table}

\begin{figure}[h]
    \centering
   \includegraphics[width=0.9\linewidth]{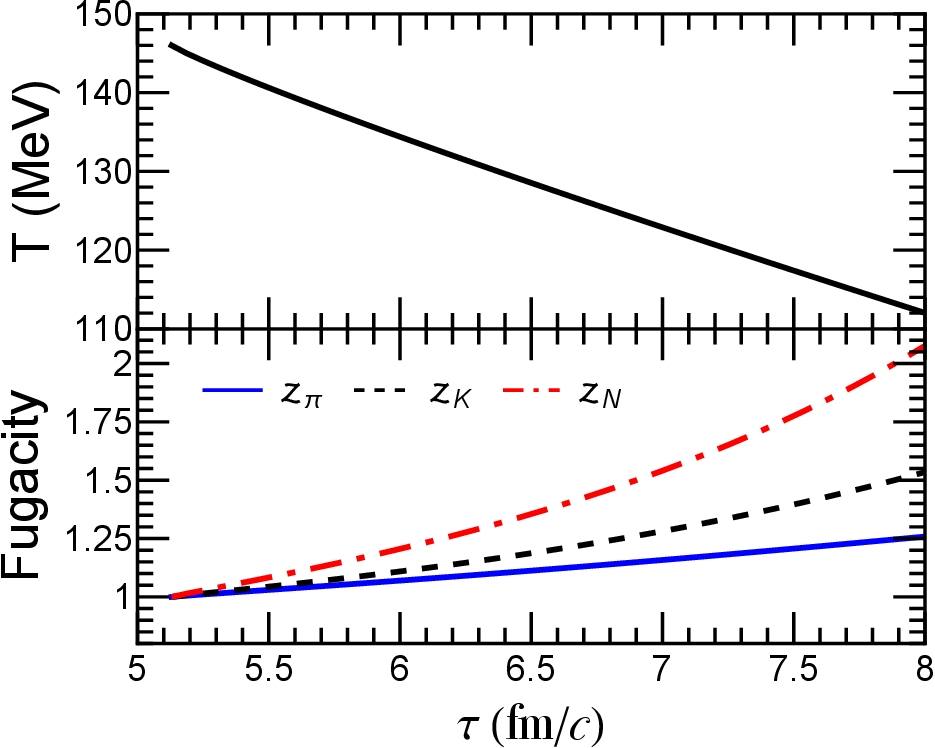}
    \caption{Time evolution of temperature (upper panel) and pion (solid line), kaon (dashed line) and nucleon (dash-dotted line) fugacities (lower panel) of the hadronic matter in Au+Au collisions at $\sqrt{s_{NN}}=7.7$ GeV.}
    \label{fig:temperature}
\end{figure}

For the time evolution of the temperature of the hadronic matter from $T_{\rm ch}=146$ MeV to $T_{\rm k}=129$ MeV, it can be obtained from the result in Ref.~\cite{Xu:2017akx}, which is based on the AMPT model simulation of Au-Au collisions at $\sqrt{s_{NN}}$=7.7 GeV, by parameterizing it as follows,
\begin{equation}\label{Temp_eq}
    T(\tau) = T_{\rm ch} -(T_{\rm ch}-T_{\rm k})\left(\frac{\tau-\tau_{\rm ch}}{\tau_{\rm k}-\tau_{\rm ch}}\right)^{0.9}.
\end{equation}
In the above, $\tau_{\rm ch}$ and $\tau_{\rm k}$ are the chemical freezeout time and the kinetic freezeout time, respectively. Values of these parameters are given in TABLE~\ref{table:temperature}. The time evolution of the hadronic matter temperature in Au-Au collisions at $\sqrt{s_{NN}}=7.7$ GeV is shown by the solid line in the upper panel of Fig.~\ref{fig:temperature}.   

For the pion fugacity $z_\pi$ in Eqs.(\ref{Eq:kinetic1}) and (\ref{Eq:kinetic2}), it can be determined together with the kaon fugacity $z_K$, and nucleon fugacity $z_N$ by requiring the constancy of the effective pion, kaon, and nucleon numbers during the expansion of the hadronic matter under the condition of constant entropy per hadron~\cite{Xu:2017akx} as described in detail in Ref.~\cite{Sung:2023oks}. Specifically, one first expresses the effective pion, kaon, and nucleon densities at temperature $T$ in terms of the thermally equilibrated hadron densities $n_i^{(0)}=g_i\int\frac{d^3{\bf p}}{(2\pi)^3}f_i({\bf p})$, where $g_i$ is the spin and isospin degeneracy of hadron species $i$. and the pion, kaon, and nucleon fugacities as follows,
\begin{eqnarray}\label{fugacity}
&&n^{\rm eff}_{\pi}(T) = z_\pi n^{(0)}_{\pi} + z_{\pi}^{2}n^{(0)}_{\rho}+z_{\pi}z_{K}n^{(0)}_{K^{*}}+z_\pi z_Nn^{(0)}_\Delta
\nonumber \\
&&~~~~~~~~~~~~+\cdots,\notag\\
&&n^{\rm eff}_{K}(T) = z_K n^{(0)}_K + z_{\pi}z_Kn^{(0)}_{K^*}+z_{\pi}^2z_{K}n^{(0)}_{K_1}+z_K^2 n^{(0)}_\phi\notag\\
&&~~~~~~~~~~~~+\cdots,\notag\\
&&n^{\rm eff}_{N}(T) = z_N n^{(0)}_N+z_\pi z_Nn^{(0)}_\Delta+\cdots,
\end{eqnarray}
where $\cdots$ denotes the contribution from strong decays of other resonances, which includes all particles of masses up to 1.7 GeV for mesons and 2 GeV for baryons in the particle data book.  In obtaining the above equations, one has used the relations $z_\rho=z_\pi^2$, $z_{K^*}=z_\pi z_K$, $z_{K_1}=z_\pi^2z_K$, $z_\Delta=z_\pi z_N$, etc. because of expected chemical equilibrium between $\pi$ and $\rho$, among $K^*$, $K$ and $\pi$, among $\Delta$, $N$ and $\pi$, etc. as a result of the large cross sections for the reactions $\pi\pi\leftrightarrow\rho$, $K^*\leftrightarrow K\pi$, $\Delta\leftrightarrow N\pi$ etc.  One then calculates the entropy density and total particle density according to $s(T)=-\sum_ig_i\int\frac{d^3{\bf p}}{(2\pi)^3}(z_if_i)\ln(z_if_i)$ and $n(T)=\sum_iz_in_i^{(0)}$. Requiring $n^{\rm eff}_{\pi,K,N}(T)V(T)=n^{\rm eff}_{\pi,K,N}(T_{\rm ch})V(T_{\rm ch})$ and $s(T)/n(T)=s(T_{\rm ch})/n(T_{\rm ch})$ and with unity pion, kaon and nucleon fugacities at $T_{\rm ch}$, these four equations can be solved to give the pion, kaon and nucleon fugacities as well as the volume ratio $V(T)/V(T_{\rm ch})$ at a given $T$.  Shown in the lower panel of Fig.~\ref{fig:temperature} are the pion fugacity $z_\pi$ (solid line), the kaon fugacity $z_K$ (dashed line) and the nucleon fugacity $z_N$ (dash-dotted line) as functions of time. One notes that the constant entropy per hadron in this study is 6.31.

\section{Results}
 
\begin{figure}[h]
    \centering
    \includegraphics[width=0.9\linewidth]{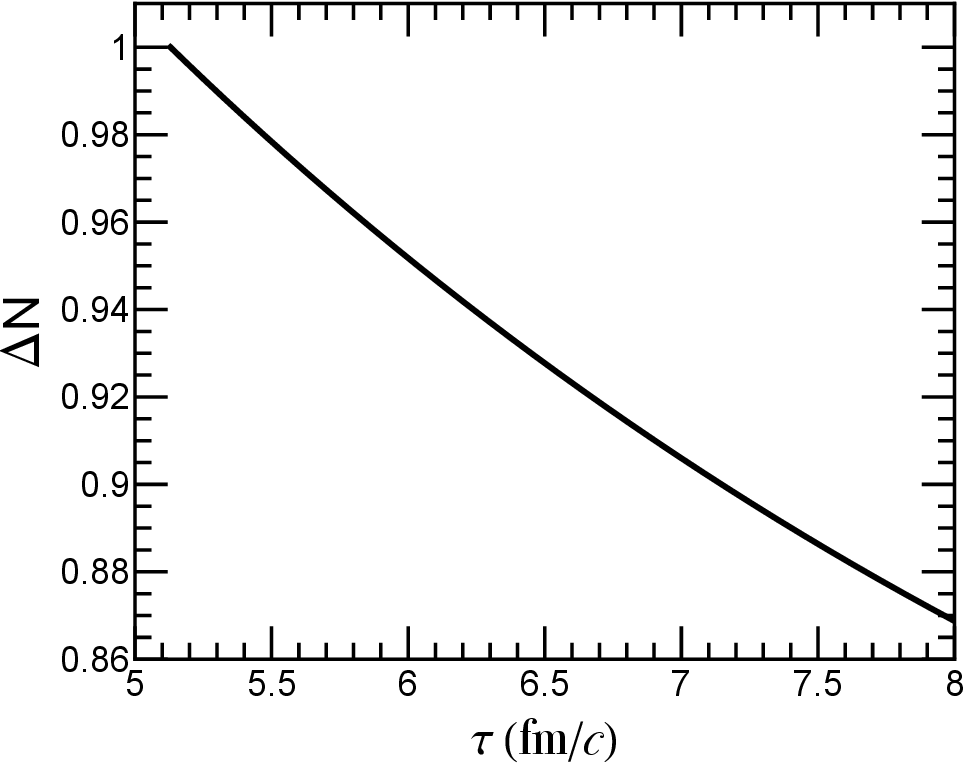}
    \caption{Time evolution of the normalized number difference between spin up and spin down $\Lambda$ hyperons in Au-Au collisions at $\sqrt{s_{NN}}=7.7$ GeV.}
    \label{fig:result}
\end{figure}

The solid line in Fig.~\ref{fig:result} shows the time evolution of the normalized number difference between spin up and spin down $\Lambda$ hyperons, which is defined by taking $\Delta N_0=1$ in Eq.(\ref{eq:number}), in Au-Au collisions at $\sqrt{s_{NN}}=7.7$ GeV. It is seen that its value decreases to 0.93 or is reduced by 7\% at the end of hadronic evolution after 1.33 fm/$c$, which justifies the neglect of hadronic scatterings on the $\Lambda$ spin polarization assumed in the literature.  The hadronic effect on $\Lambda$ polarization remains small even one doubles the lifetime of the hadronic phase in these collisions to 2.66 fm/$c$ as the value of $\Delta N$ is now 0.88 or is reduced by 12\% in this case.  

\section{Conclusions and discussions}

The effects of hadronic scatterings on the $\Lambda$ spin polarization in Au-Au collisions at $\sqrt{s_{NN}}=7.7$ GeV are studied in a schematic kinetic approach by using thermally averaged $\Lambda$ spin flip and non-flip cross sections. These cross sections are calculated by including contributions from both the $s$-channel process through the $\Sigma^*$ resonance and the $t$-channel process via the exchange a scalar $\sigma$ meson. Neglecting the contribution from the $t-$channel process because of its negligible small ratio of spin flip to non-flip cross sections compared to that in the $s-$channel $\Lambda-\pi$ resonance scattering process, which has a constant value of 1/3.5, the $\Lambda$ spin polarization is found to decrease by only about 7-12\% during the hadronic stage of these collisions. Results from the present study thus justify the assumption in theoretical studies of $\Lambda$ polarization that compare its value calculated at the chemical freezeout to the measured one at the kinetic freezeout.  

The present study has also neglected the interference between the $s-$channel and $t-$channel $\Lambda-\pi$ scattering processes. Because of the negligible spin flip to non-flip cross sections in the $t$-channel process, including this term is not expected to affect the results and conclusion from the present study.   Further neglected in the present study is the effect of $\Lambda$ scattering by nucleons, whose number in the hadronic matter becomes non-negligible in collisions at $\sqrt{s_{NN}}=7.07$ GeV~\cite{STAR:2017ckg} due to partial stopping of the colliding nuclei.  The $\Lambda-N$ spin flip to non-flip scattering ratio due to the $\sigma$ meson exchange is, however, also expected to be negligibly small as in $\Lambda-\pi$ scattering because of the small $\Lambda$ velocity, which makes the spin flip effect negligible. Although $\Lambda-\pi$ and $\Lambda-N$ scatterings can also go through the exchange of an $\omega$ meson, these scatterings again will not affect much the  polarization of a $\Lambda$ hyperon due to its small velocity.  Therefore, the conclusion reached in the present study, i.e., hadronic scatterings can hardly affect the $\Lambda$ polarization in relativistic heavy ion collisions, remains unchanged after all these scattering processes are included in the kinetic equations for the evolution of the $\Lambda$ polarization during the expansion and cooling of the hadronic matter.

However, the present study has only considered hadronic scattering effects on the $\Lambda$ polarization. Since the measured $\Lambda$ polarization in relativistic heavy ion collisions is the weighted average of the polarizations of primary $\Lambda$ and the $\Lambda$ from the radiative decay of $\Sigma^0$ or the strong decay of strange baryon resonances, particularly the $\Sigma^*$ resonance, at the kinetic freezeout of these collisions~\cite{Becattini:2019ntv,Xia:2019fjf}, hadronic scattering effects on the polarizations of $\Sigma^0$ and $\Sigma^*$, besides that on the $\Lambda$, need to be included for a more complete study. This requires treating $\Sigma^0$ and $\Sigma^*$ explicitly in the kinetic equations and following the time evolution of their numbers in various spin states, instead implicitly for the $\Sigma^*$ in the present study through the $\Lambda-\pi$ resonance scattering.   Due to the finite lifetime of $\Sigma^*$, the $\Lambda-\pi$ resonance scattering in this coupled spin kinetic approach becomes a two-step process of $\Lambda+\pi\to\Sigma^*$ and $\Sigma^*\to\Lambda+\pi$ with the $\Sigma^*$ lasting about $\Gamma_{\Sigma^*}^{-1}=5.5~{\rm fm}/c$ between creation and decay.  The latter is expected to result in an even smaller decrease of the $\Lambda$ polarization during the hadronic expansion than in the present study of treating the $\Lambda-\pi$ scattering instantaneously in time.  On the other hand, the feed-down contribution to $\Lambda$ polarization from $\Sigma^*$ strong decay might be reduced due to possible suppression of its production as for the production of other hadron resonances in relativistic heavy ion collisions~\cite{Cho:2015qca,Knospe:2021jgt}.  Studies based on such a coupled spin kinetic approach will be of great interest for a quantitative understanding of the hadronic scattering effects on $\Lambda$ polarization in relativistic heavy ion collisions. 

\section*{ACKNOWLEDGEMENTS}

This work was supported by the Korea National Research Foundation under Grant No. 2023R1A2C300302311, 2023K2A9A1A0609492411 (S.H.L.) and the U.S. Department of Energy under Award No. DE-SC0015266 (C.M.K.). H. Sung thanks the Cyclotron Institute of Texas A\&M University for its hospitality during her stay as a visiting scholar.

\bibliography{main.bib}
\end{document}